# Delivering 10 Gb/s optical data with picosecond timing uncertainty over 75 km distance

N. Sotiropoulos,[1] C. M. Okonkwo,[1] R. Nuijts,[2,4] H. de Waardt,[1] and J. C. J. Koelemeij[3,*]

[1] *Department of Electrical Engineering, COBRA Research Institute, Eindhoven University of Technology, Den Dolech 2, P.O. Box 513, NL 5600 MB Eindhoven, The Netherlands*
[2] *Network Services, SURFnet, Radboudkwartier 273, P.O. Box 19035, 3501 DA Utrecht, The Netherlands*
[3] *LaserLaB, Department of Physics and Astronomy, VU University, De Boelelaan 1081, 1081 HV Amsterdam, The Netherlands*
[4] *Currently with Ciena Corporation, Ferring Building, Polaris Avenue 140A, 2132 JX Hoofddorp, The Netherlands*
[*] *j.c.j.koelemeij@vu.nl*

**Abstract:** We report a method to determine propagation delays of optical 10 Gb/s data traveling through a 75 km long amplified fiber link with an uncertainty of 4 ps. The one-way propagation delay is determined by two-way exchange and cross correlation of short (< 1 ms) bursts of 10 Gb/s data, with a single-shot time resolution better than 2.5 ps. We thus achieve a novel optical communications link suited for both long-haul high-capacity data transfer and time transfer with picosecond-range uncertainty. This opens up the perspective of synchronized optical telecommunication networks allowing picosecond-range time distribution and millimeter-range positioning.

## 1. Introduction

Accurate time distribution and clock synchronization have become essential ingredients of many network technologies vital to our society, examples including satellite navigation systems, telecommunication networks, electrical power grids, and networks for financial trading [1]. Clock synchronization in such networks is often achieved through satellite-based methods such as global navigation satellite systems (GNSS) [1], and the most accurate techniques yield a time uncertainty slightly below 1 ns [2,3]. Currently, fiber-optic methods for long-distance time and frequency transfer are being widely investigated as an alternative to satellite-based methods. Recent state-of-the-art results include optical frequency transfer through installed fiber links with lengths up to 1840 km [4,5], and two-way time transfer (TWTT) through installed fiber links up to 540 km length [5-7]. During TWTT, two clocks are synchronized through the exchange of electromagnetic pulses while correcting for the one-way delays of these pulses. Estimating and correcting for propagation delays is an important aspect of any time distribution system and underlies the principle of operation of *e.g.* GNSS. Recent work on fiber-optical TWTT has resulted in observed clock offsets as small as 6 ps (with a standard uncertainty $\sigma = 8$ ps) after synchronization through a link of 62 km length, and 35 ps ($\sigma = 20$ ps) for a link of 480 km length [7]. The uncertainty in the clock offsets is primarily due to the estimated value of the one-way delay (OWD), which is determined from measured signal round trip delays (RTDs). To our knowledge, the values reported in Ref. [7] represent the most accurate estimates of one-way signal propagation delays over distances in the 50 – 500 km range to date. Apart from absolute time accuracy, an important figure of merit of any TWTT system is the time stability (TDEV), which is a measure of the averaging time needed to achieve a certain synchronization level [8]. For TWTT through the 62 km (480 km) link in Ref. [7] this equals about 2 ps after 100 s averaging, reaching a minimum of 0.3 ps (0.6 ps) after $7\times10^4$ s ($3\times10^3$ s). Other groups have reported time accuracies in the range 74 – 250 ps, and observed time stabilities of 3 – 5 ps after 200 – 300 s averaging [5,6]. Typically, these methods make use of relatively low-bandwidth signals (radiofrequency domain) modulated onto an optical carrier, which is transmitted through 'dark fiber' (a fiber in which no other optical signals are present) [6,7], or through a 'dark channel' (a certain amount of bandwidth –typically 50 GHz or more– reserved exclusively for time and frequency signals) [5]. In such methods, and contrary to the original purpose of installed fiber, no data other than time and/or frequency information are transmitted. As a result, fiber-optic TWTT requires sacrificing valuable optical telecommunication bandwidth, and obtaining access to installed fiber for scientific TWTT has typically been either expensive, or restricted to fiber links operated by national research and education networks. A natural solution to circumvent this would be to integrate the functionalities of optical telecommunication and high-accuracy time transfer into a shared data modulation format. Several approaches in this direction have been investigated, and time stabilities well below 1 ns have been demonstrated [9-11]. However, fully integrated TWTT and data transfer has so far only been achieved by so-called White Rabbit Ethernet, based on refined existing protocols for joint data and time transfer (IEEE 1588v2 and Synchronous Ethernet) and enabling 1 Gb/s data transfer and TWTT with a time uncertainty well below 1 ns over fiber links up to 10 km [12,13].

Here, we demonstrate a method to determine propagation delays (as required for TWTT) of error-free 10 Gb/s optical bit streams through a 75 km amplified fiber link, achieving an unprecedented single-shot measurement uncertainty of 4 ps while using no other signals than the 10 Gb/s bit streams themselves. With a single shot lasting less than 1 ms, the method also offers high stability. In Sec. 2, the principle of operation of our method is described. In Sec. 3 we present the results and limitations of our method, and we propose an advantageous strategy to implement the method in installed, live fiber links for optical telecommunication. Results are presented in Sec. 3. Conclusions are presented in Sec. 4, together with an outlook describing the potential of our method for future terrestrial timing and positioning systems.

## 2. Methods

### 2.1 Fiber-optical link design and optical modulation format

The basic principle underlying our method consists of a measurement of the RTD of the fiber link, which is divided by two to find the OWD, after correcting for delay asymmetries present in the link. We determine delays by cross correlation of optical data streams in the time domain, similar to the delay calibration by cross correlation of pseudo-random bit sequences (PRBSs) employed in satellite time transfer and positioning. Our setup consists of components typically found in commercial optical telecommunications systems, and is depicted schematically in Fig. 1(a). Optical 10 Gb/s data streams are transferred between two locations A (the location of the reference clock) and B (the remote location where, for example, a second 'slave' clock needs be synchronized with the reference clock), connected by up to 75 km of fiber length. A 10 Gb/s pattern generator [Fig. 1(a)] is used in location A to drive an electrical-to-optical (E/O) converter consisting of a Mach-Zehnder modulator with a chirp parameter $\alpha = 0.2$-$0.3$, which modulates the amplitude of an optical carrier (narrow-band laser source with wavelength $\lambda_1$=1552.52 nm/ITU channel #31) employing on-off keying (OOK). The optical signal is multiplexed into a first fiber spool of 25 km length. To overcome the signal attenuation due to link losses, and to demonstrate the suitability of the method for amplified links, the optical signal is demultiplexed and sent through a quasi-bidirectional optical line amplifier (OLA) [14]. The quasi-bidirectional amplifier is built up from two semiconductor optical amplifiers (SOAs) equipped with optical isolators (OIs), which limit the influence of backscattered light on the SOAs, thereby allowing operation at high gain. After amplification by the first SOA of the OLA, the optical signal is multiplexed into either a short (few meter) fiber link, or one or two 25-km fiber spools connected in series. This is done using wavelength multiplexers (WMs) based on arrayed waveguide gratings (100 or 200 GHz channel spacing). In this way, fiber links of 25, 50 and 75 km length are realized, with a maximum SOA gain of 15 dB for the longest link. At the remote end (location B), the signal is demultiplexed and fed into a 10 Gb/s receiver consisting of a photodetector in conjunction with a transimpedance amplifier and a limiting amplifier, which converts the optical OOK signal into an electrical signal. The receiver electrical output contains the original PRBS delayed by exactly the OWD (denoted as $t_{AB}$) by the time it reaches location B in the setup. The limiting amplifier has two output ports, and the first port is connected either to the input of a bit-error-rate tester (BERT), or to one of the four input channels of a 50 GS/s real time digital phosphor oscilloscope (DPO) that is referenced to an external 10 MHz rubidium (Rb) atomic clock. In practical applications, this clock may also act as the reference clock to which other (remote) clocks in a network must be synchronized. The signal from the second output port of the limiting amplifier at B is used to drive a second E/O converter, which modulates an optical carrier at $\lambda_2$=1550.92 nm (ITU channel #33) through OOK. This optical signal is multiplexed into the fiber link, thus sending the optical signal back to location C (which in our setup is co-located with A) via essentially the same path except for the quasi-bidirectional OLA, where the optical signal is routed through the second SOA (using wavelength multiplexers) before being fed back into the first fiber spool. Once arrived at C, a second 10 Gb/s receiver converts the optical OOK signal to an electrical signal. Again, the receiver output contains the original PRBS, now delayed by an amount $t_{AC} \equiv t_{AB} + t_{BC}$. This signal is fed into a second channel of the DPO, while a third channel captures the original (undelayed) PRBS produced by the pattern generator.

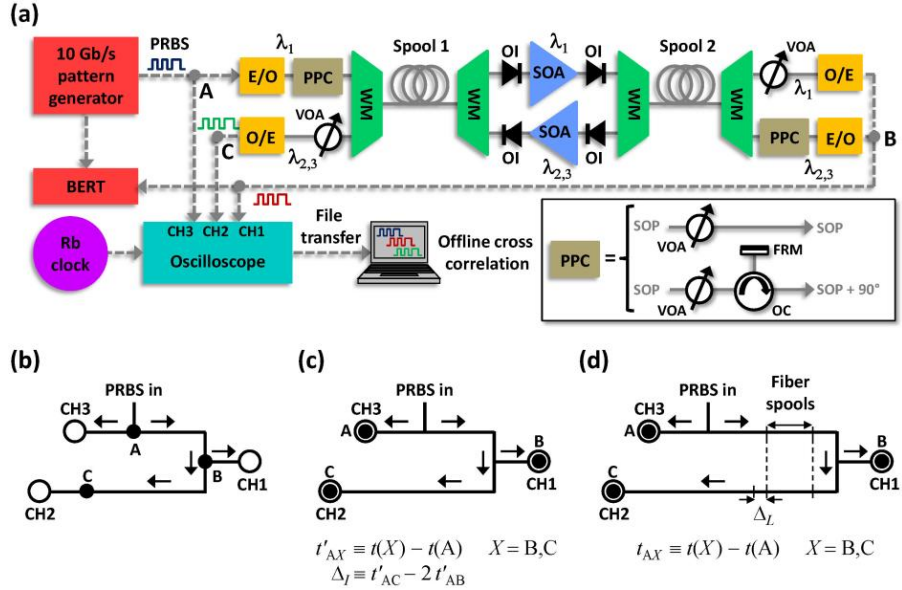

Fig. 1. (a) Experimental setup for 10 Gb/s data transmission and picosecond-uncertainty delay measurements over 75 km distance. Solid lines represent optical connections, whereas electrical connections are indicated by dashed lines. E/O, electrical-to-optical converter; O/E optical-to-electrical converter; BERT, bit-error-rate tester; WM, wavelength multiplexer; SOA, semiconductor optical amplifier; OI, optical isolator; VOA, variable optical attenuator. Both the upstream and downstream channels are equipped with a polarization and power control (PPC) unit (inset). The PPC contains a VOA, and can furthermore be configured to rotate the state of polarization (SOP) by either 0° or 90° by use of an optical circulator (OC) and a Faraday rotator mirror (FRM). The VOAs and PPCs allow maintaining a constant optical power level in the system as well as the determination of PMD. (b) Schematic of the signal propagation directions and delays in the setup shown in (a) with the fiber spools removed. (c) Same as in (b), but with the reference points A, B and C now defined such that they coincide with the DPO channels 3, 1 and 2, respectively. In the definition of the instrument delays $t'_{AX}$, $t(X)$ denotes the point in time at which the PRBS arrives at location $X$. (d) Signal propagation delays with the fiber spools inserted in the link. The link delay asymmetry is modeled by an additional delay $\Delta_L$ in the return path.

*2.2 Cross correlation method*

The 10 Gb/s PRBS signals at the three DPO channels are captured at 12.5 GS/s during 1 ms [Fig. 2(b)]. After capture, the signal records are saved to a removable disk (which takes about three minutes for the three 0.45 GB signal records), and subsequently transferred to a computer where a software algorithm performs the cross correlation in order to infer the delays $t_{AB}$ and $t_{AC}$. The time delay between any two PRBSs is found by determining the peak position of their cross-correlation spectrum [blue dots in Figs. 2(c) and 2(d)]. Given the 12.5

GS/s sampling rate, the precision is limited to about 80 ps. To enhance the precision, we fit a high-resolution model spectrum to the discrete 12.5 GS/s correlation feature using a non-linear least-squares fitting algorithm. The model spectrum is obtained by interpolation of a correlation spectrum of 10 Gb/s data of shorter duration, but captured at a higher sampling speed (50 GS/s). The latter data were acquired using a shifted copy of the reference PRBS, produced by a second output channel of the pattern generator [not shown in Fig. 1(a)]. This offers higher resolution at the expense of absolute time accuracy, since the constant delay caused by the shift is not well calibrated. As shown in Figs. 2(c) and 2(d), the fit method employed here offers both high resolution (we here define the resolution as the fit standard error, which is typically between 1 and 2.5 ps) and high time base accuracy, in spite of the fact that the PRBS is under-sampled at 80 ps-per-point intervals. For the 75 km link, $t_{AC} \approx 2\, t_{AB} \approx$ 0.74 ms. The 10 Gb/s PRBS word length is set to $2^{23} - 1$ symbols so that the PRBS repeats itself only once every 0.84 ms. This prevents the appearance of aliases in the correlation spectrum, although these could be identified with the aid of a coarse estimate of the link delay by other means (*e.g.* from length and refractive index specifications by the fiber manufacturer).

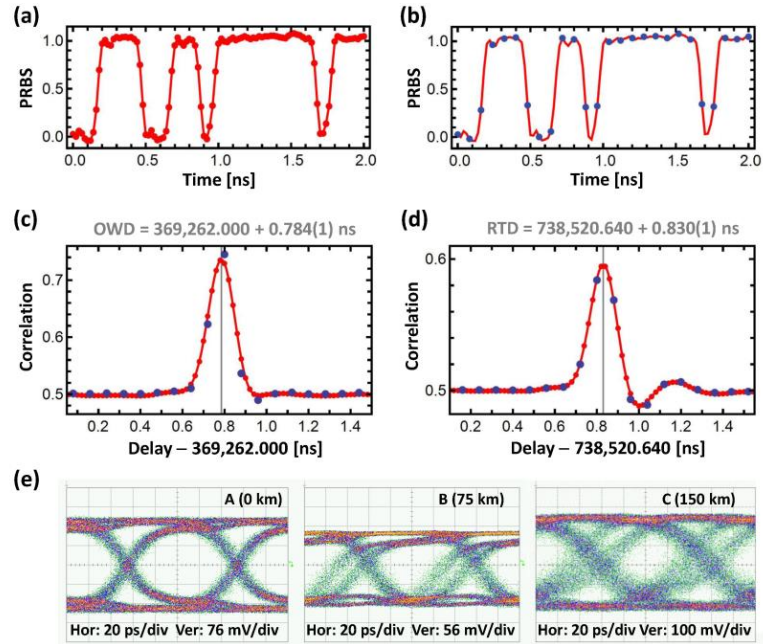

Fig. 2. Delay determination by cross correlation of PRBSs. PRBS captured by the real-time DPO in location A [Fig. 1(a)] at sampling rates of 50 GS/s [red dots, Fig. 2(a)] and 12.5 GS/s [blue dots, Fig. 2(b)], respectively. Red curves are interpolations of the 50 GS/s data and serve to guide the eye. Similar signals are captured at channels 1 and 2 of the real-time DPO (which correspond to locations B and C, respectively). (c) Normalized cross correlation spectrum of the PRBS signals at locations A and B. Blue dots represent the correlation spectrum of the PRBSs captured at 12.5 GS/s with respect to the Rb-clock-stabilized time base of the DPO. Red dots are the high-resolution (50 GS/s) data obtained with the delayed-reference method (see Sec. 2.2), and are shown together with their interpolating function which is used to fit the blue data points and determine the peak position (indicated by the vertical grey line), which yields a direct reference determination of the OWD. The delay offset in the abscissa label arises during the analysis and has no further physical meaning. (d) Cross correlation spectrum of the PRBS signals at locations A and C, and the fitted peak position, which corresponds to the RTD. (e) Eye diagrams captured with a sampling oscilloscope of the PRBS signals at locations A, B and C, respectively. For data transmission, only the signals in A and B are relevant.

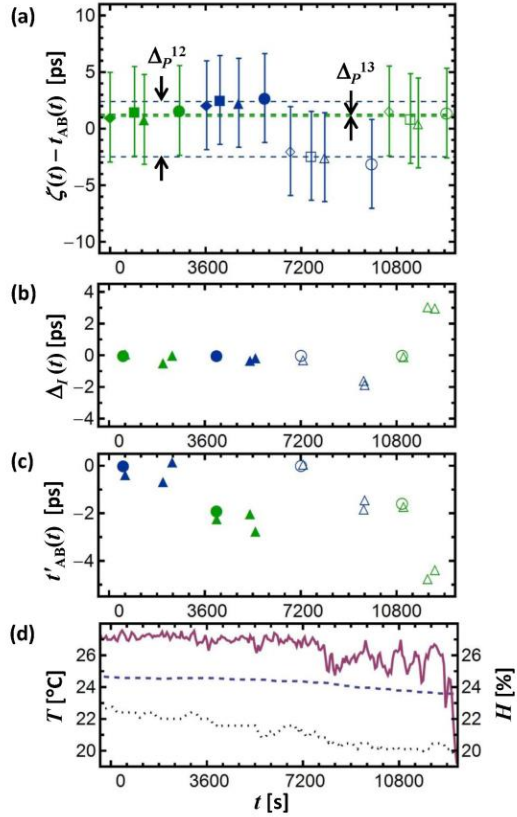

Fig. 3. (a). Time differences $\zeta(t) - t_{AB}(t)$ obtained for the 75 km link at different times $t$ during one day of measurements. Green and blue data points are obtained using wavelength combinations $(\lambda_1, \lambda_3)$ and $(\lambda_1, \lambda_2)$, respectively. The delay asymmetry due to dispersion is removed using Eqs. (1) and (2), assuming the value for the chromatic dispersion found from Eq. (3) after PMD correction. Filled symbols indicate the use of FRMs in the PPCs [Fig. 1(a)], whereas open symbols correspond to data acquired without FRMs. Horizontal dashed lines represent the mean of each group of four data points; differential delays $\Delta_P^{1j}$ are derived from the difference of these means as indicated by the arrows. Pairs of data points with identical symbols (but with opposite wavelength combinations) correspond to data which are combined using Eq. (4) to obtain the results of Fig. 4(d). (b) Variation of $\Delta_I$ during the measurements. Only the data corresponding to the circles/disks are used for calibration; triangle data are used for drift monitoring only and are relative to the first data point. Symbol filling and color correspond to that used in (a). (c) Variation of $t'_{AB}$ during the measurements with respect to $t'_{AB}(t=0\ s)$ (first eight data points) and $t'_{AB}(t=7230\ s)$ (final eight data points). (d) Temperature $T$ and relative humidity $H$ during the measurements: solid curve, $T$ measured at the O/E converter at location B; dashed curve, $T$ measured inside spool housing; dotted curve, $H$ measured at the optical workbench. The steep temperature drop near $t=12500$ s is due to the laboratory heating system being switched off at 5 pm local time.

The accuracy of the measured time delays is fundamentally limited by the instability of the DPO time base with respect to the Rb reference clock, which we quantify as follows. We split off half of the 10 MHz reference signal of the Rb clock, and record it using the fourth channel of the DPO. From the signal record we retrieve the TDEV with respect to the phase of the Rb clock [8], which we find to equal 2.4 ps (0.3 ps) for 0.1 ms (1 ms) averaging time.

*2.3 Determination of delays and delay asymmetry*

The choice of the reference points A, B and C in Figs. 1(a) and 1(b) is helpful for the qualitative description of the signal paths and propagation delays in Sec. 2.1. However, for the analysis of the delays obtained through cross correlation it is convenient to define A, B and C such that they coincide with channels 3, 1, and 2 of the DPO, respectively. More precisely, the points A, B and C coincide with the points where the sampling of the signals occurs with respect to the DPO time base [Figs. 1(c), 1(d)]. Thus, in what follows, the definition of A, B and C shown in Figs. 1(c) and 1(d) will be used.

In contrast to our laboratory setup, in a long-haul transmission link only $t_{AC}$ can be accurately measured as the clock at the remote location B is not (yet) synchronized. The aim therefore is to infer the OWD $t_{AB}$ from $t_{AC}$, which requires that the delay asymmetry $\Delta \equiv t_{BC} - t_{AB}$ be known so that the OWD can be determined from the relation $t_{AB} = (t_{AC} - \Delta)/2$. In our setup, we distinguish between the instrument delay asymmetry, $\Delta_I$ [Fig. 1(c)], due to the nonreciprocal paths in the transceiver systems at locations A and B and the quasi-bidirectional amplifier, and the link delay asymmetry, $\Delta_L$ [Fig. 1(d)], associated with dispersion in the bidirectional optical fiber link. We determine $\Delta_I$ (which is of the order of 20 ns) by removing the fiber spools and measuring the instrument delays $t'_{AB}$ and $t'_{AC}$ via cross correlation of DPO traces captured at 50 GS/s [Fig. 2(a)]. Hence, $\Delta_I = t'_{AC} - 2t'_{AB}$ [Fig. 1(c)]. During this procedure, we adjust four variable optical attenuators (VOAs) so as to keep the optical power at the SOAs and receivers constant to within 0.3 dB [Fig. 1(a)]. This introduces a possible parasitic differential delay in the VOAs themselves which we constrain to 0(1) ps. We observe that instrument delays may vary by many ps over the course of hours, as is illustrated in Fig. 3(c). Such variations are fully accounted for by our method as long as $\Delta_I$ stays constant. We have therefore studied the dependence of $\Delta_I$ on a number of environmental and link conditions which vary over time (temperature, humidity, received optical power, plugging/unplugging fiber connections), the details of which will be published elsewhere. We find that $\Delta_I$ can be trusted to within 3 ps at the ~30 minutes time scale of the measurements and for typical environmental conditions in our laboratory [Figs. 3(b), 3(d)]. As is often the case in conventional optical communications links, no special measures are taken to reduce acoustical noise or temperature and humidity variations in the setup. Here it is important to mention that the temperature of the SOAs is controlled to within 0.1 K using thermo-electric coolers, so that a variation of the ambient temperature by 10 K has no visible effect on $\Delta_I$ at the 0.3 ps level (limited by the precision of our method). By contrast, if the temperature difference between the SOAs is allowed to increase to the 1 K level, changes in $\Delta_I$ at the level of 5-10 ps are observed. The link delay asymmetry ($\Delta_L$) is caused primarily by the different propagation speeds of the wavelengths used for downstream (direction A to B) and upstream communication due to material, waveguide and polarization mode dispersion (PMD). Other possible causes are fast random optical path length fluctuations of acoustic origin which might take place within the 0.74 ms round-trip time. Noise measurements performed in fiber spools of similar length point out that the latter effect contributes far less than 0.1 ps to $\Delta_L$ [15], and we do not consider this further here. Two other sources of delay asymmetry are self-phase modulation (SPM) and cross-phase modulation (XPM), which are caused by refractive-index changes due to the presence of optical fields [16]. We estimate these two effects to contribute less than 0.1 ps and we ignore them here. Material and waveguide dispersion, however, have a much larger effect. Starting out from the well-known expression for the group delay, $\tau(\lambda) = [n(\lambda) - \lambda n'(\lambda)]L/c$, of a modulated optical signal travelling through a dielectric medium with length $L$, effective index of refraction $n(\lambda)$, and with $c$ the speed of light, we find the delay asymmetry due to chromatic dispersion as

$$\tau(\lambda) - \tau(\lambda_1) = -\frac{L}{2c}\left[2(\lambda - \lambda_1)\lambda_1 n''(\lambda_1) + (\lambda - \lambda_1)^2 \left(n''(\lambda_1) + \lambda_1 n'''(\lambda_1)\right)\right]. \tag{1}$$

Here we neglect terms of order $(\lambda - \lambda_1)^3$ and higher. Note that this expression does not take into account PMD, which we determine independently as explained further below. It is possible to determine $n''(\lambda)$ and $n'''(\lambda)$ from a measurement of the dispersion, $D(\lambda) = -\lambda\, n''(\lambda)/c$, by detecting AM power variations of an optical signal transmitted through the fiber link as the AM frequency is swept from 0 to 20 GHz [17]. We use this method to find a dispersion value

$D = 16.5(1)$ ps/nm km, which implies about 2 ns of delay asymmetry for a 75 km link. To find $n'''(\lambda)$ we use the dispersion formula $D(\lambda) = (S_0/4)(\lambda - \lambda_0^4/\lambda^3)$, with $\lambda_0=1.31(1)$ μm and $S_0$ adjusted so as to match the measured dispersion in our fiber. We thus find that the term in $n'''(\lambda)$ in Eq. (1) adds significantly (4 ps) to the delay asymmetry to a round trip over a 75 km link. With $n''(\lambda)$ and $n'''(\lambda)$ known, Eq. (1) may be used to correct for the delay asymmetry due to dispersion, so that the estimated OWD, $\zeta^{1j}$, obtained from a round trip using an downstream wavelength $\lambda_1$ and upstream wavelength $\lambda_j$, becomes

$$\zeta^{1j} = \frac{1}{2}\left(t_{AC}^{1j} + \tau(\lambda_1) - \tau(\lambda_j) - \Delta^{1j}\right), \tag{2}$$

with $\tau(\lambda_1) - \tau(\lambda_j)$ given by Eq. (1), and where $\Delta^{1j}$ may account for all possible delay asymmetries (due to *e.g.* PMD) other than those contained within Eq. (1). However, the measurement uncertainty in $D$ translates to a delay uncertainty of 12 ps in Eq. (2). Moreover, the dispersion may vary over time due to environmental conditions, so that the dispersion would have to be re-measured periodically. To circumvent the need for an independent dispersion determination altogether, we add a third optical wavelength channel, $\lambda_3=1549.32$ nm (ITU channel #35) for communication from location B to C. Thus two RTDs, $t_{AC}^{12}$ and $t_{AC}^{13}$, are measured, employing the wavelength combinations $(\lambda_1, \lambda_2)$ and $(\lambda_1, \lambda_3)$, respectively. We also measure the corresponding instrument delays $t'_{AC}^{12}$ and $t'_{AC}^{13}$. This information can be combined with the theoretically expected RTDs based on the expression for $\tau(\lambda)$, from which the dispersion (and therefore $n''(\lambda)$) is retrieved as (again neglecting terms of order $(\lambda-\lambda_1)^3$ and higher)

$$\begin{aligned}D(\lambda_1) &= -\frac{\lambda_1}{c}n''(\lambda_1) \\ &= \frac{2\lambda_1}{L(\lambda_2^2 - \lambda_3^2)}\left\{\left(t_{AC}^{12} - \Delta^{12}\right) - \left(t_{AC}^{13} - \Delta^{13}\right)\right\} + \frac{\lambda_1^2(\lambda_2 + \lambda_3 - 2\lambda_1)n'''(\lambda_1)}{c(\lambda_2 + \lambda_3)}.\end{aligned} \tag{3}$$

During the experiments, all wavelengths are measured intermittently using a wavelength meter with 0.3 pm measurement uncertainty. The three wavelengths vary about their mean value in a correlated fashion, with a standard deviation of at most 1.5 pm.

Employing the same approximations as used in the derivation of Eq. (1), we combine the expression for $\tau(\lambda)$ with Eqs. (1) and (3), and after some algebraic manipulation we derive the estimate of the OWD, $\theta_{AB}$,

$$\theta_{AB} = \frac{1}{2(\lambda_2^2 - \lambda_3^2)}\Big[\left\{\left(t_{AC}^{12} - \Delta^{12}\right) - \left(t_{AC}^{13} - \Delta^{13}\right)\right\}\lambda_1^2 + \left(t_{AC}^{13} - \Delta^{13}\right)\lambda_2^2 - \left(t_{AC}^{12} - \Delta^{12}\right)\lambda_3^2 \\ - L\lambda_1^2(\lambda_1 - \lambda_2)(\lambda_1 - \lambda_3)(\lambda_2 - \lambda_3)n'''(\lambda_1)/c\Big]. \tag{4}$$

In what follows, we will use the definition $\Delta^{1j} \equiv \Delta_I^{1j} + \Delta_P^{1j}$, with $\Delta_P^{1j}$ the delay asymmetry due to PMD. We determine the $\Delta_P^{1j}$ by performing half of the measurements with an optical circulator and Faraday mirror installed behind each E/O converter, and the other half without these components installed [Fig. 1(a)]. In this way, the input state of polarization (SOP) of both the downstream and upstream signals is rotated by 90°, which inverts the sign of the PMD delay asymmetry and allows determining $\Delta_P^{12}$ and $\Delta_P^{13}$ [Fig 3(a)]. We observe that the $\Delta_P^{1j}$ are constant throughout the total measurement duration, and the largest observed delay asymmetry due to PMD was 4.9(6) ps (for the 75 km link). In addition, we measure the differential group delay due to PMD independently with a polarization analyzer, which confirms that delay asymmetries of up to 6 ps may occur in the 75 km link. As PMD in optical fiber is typically due to birefringence which varies along the fiber in a random fashion, the sign and magnitude of differential delays vary randomly with wavelength [16]. Measurements with the polarization analyzer confirm the presence of such behavior in our fiber link, which is also reflected by Fig. 3(a), where the value of $\Delta_P{}^{12}$ is nearly 5 ps, whereas $\Delta_P{}^{13}$ is close to 0 ps.

Equation (4) assumes that $t_{AC}{}^{12}$ and $t_{AC}{}^{13}$ are obtained simultaneously, which requires that the two upstream wavelength channels ($\lambda_2$, $\lambda_3$) are operated simultaneously. However, to minimize the complexity of our setup we choose to measure $t_{AC}{}^{12}$ and $t_{AC}{}^{13}$ at times separated by 10 – 60 minutes, during which we exchange the wavelength sources ($\lambda_2$, $\lambda_3$) and re-configure the connections to the WMs accordingly. During this period the delay associated with the reciprocal part of the fiber link, $t_{AB} - t'_{AB}$, may change by 0.1 – 1 ns primarily due to temperature-induced optical path length variations. We correct for these delay changes by making the replacement $(t_{AC}{}^{13} - t'_{AC}{}^{13}) \rightarrow (t_{AC}{}^{13} - t'_{AC}{}^{13}) \times (t_{AB}{}^{12} - t'_{AB}{}^{12})/(t_{AB}{}^{13} - t'_{AB}{}^{13})$ in Eq. (4). We thus achieve a quasi-simultaneous measurement of $t_{AC}{}^{12}$ and $t_{AC}{}^{13}$ with less than 0.2 ps additional measurement error (primarily due to possible correlated path length changes in instrument leads, fiber patches, *etc.*, which are ignored here).

## 3. Results and discussion

The uncertainties of all input parameters can be propagated through Eq. (4) to find the uncertainty of $\theta_{AB}$, yielding a total uncertainty of 4 ps. Table 1 gives an overview of the various contributions to this uncertainty. We assess the validity of our estimate $\theta_{AB}$ by comparing the values of eight OWD determinations $\theta_{AB}$ with the directly measured values of the OWD, $t_{AB}$. We have followed this procedure using optical links of 25, 50 km and 75 km of length, and the results are shown in Figs. 4(b)-4(d). For example, for the first data point in Fig. 4(d), $\theta_{AB}$ = 369,287,563.9(4.2) ps, and $t_{AB}$ = 369,287,565.6(0.8) ps. All 24 delay differences ($\theta_{AB} - t_{AB}$) are within ±5 ps, while 20 are within ±4 ps, in good agreement with the estimated uncertainty. For each set of delay differences (for the 25 km, 50 km and 75 km links), we compute the mean and standard deviation (given within parentheses) to find 0.0(0.7) ps, 3.1(1.5) ps, and −1.0(0.8) ps, respectively. For the 50 km and 75 km links a statistically significant offset is observed, which suggests that systematic effects have a larger impact on the measurement uncertainty than measurement noise. However, we find no sign of systematically increasing timing errors as the link becomes longer, indicating that length-dependent delay asymmetries are taken properly into account. We furthermore point out that each delay determination requires only a single shot of data, lasting less than 1 ms. This suggests that extremely fast picosecond time transfer may be possible with our method, *i.e.* with orders of magnitude higher stability than existing state-of-the-art methods [5-7,18], provided that both the data acquisition and cross correlation algorithm are implemented in hardware so that signals can be processed in real time. From our results we can also derive an accurate value of the fiber dispersion. Using Eq. (3), and after correcting for delays due to PMD, we find dispersion values of 16.56(4) ps/(nm km), 16.55(3) ps/(nm km), and 16.52(4) ps/(nm km) for the 25 km, 50 km, and 75 km links, respectively.

The uncertainty budget in Table 1 reveals that the accuracy is limited largely by the calibration of delay asymmetries in the instruments. Other sources of uncertainty include the resolution of the cross correlation methods, the DPO time base stability, and the uncertainty of parameter values which determine the delay asymmetry due to dispersion in the link. For the 25 km, 50 km and 75 km links we furthermore verify that the 10 Gb/s data stream is detected in location B with essentially zero bit errors using a bit-error-rate tester [Fig. 4(a)]. Therefore, when deployed in existing, installed long-haul fiber links, our method enables both high-capacity data transfer at 10 Gbps and picosecond-level synchronization of clocks in a network. This requires that all instrument delay asymmetries be calibrated, either prior to installation, or through some automated local calibration routine that uses a time base derived from either a simple stand-alone local oscillator, or the not-yet-synchronized clock, or the recovered clock signal obtained from the optical data stream. We observe that if the setup remains untouched while temperature variations are kept below one Kelvin, the instrument delay asymmetry remains constant to within one picosecond over the course of five hours. We therefore anticipate that with simple thermal control, instrument delay asymmetries may be kept constant to within a few picoseconds at all times, and the long-term stability of the fiber link will be subject of future study. Kinematic effects such as the Sagnac effect due to the rotation of the Earth and relativistic effects (which vanish for our setup as A and B are co-located) must be taken into account as well [5,7,19].

Table 1. Uncertainty budget for $\theta_{AB}$

| $L$ [m] | 25 255(1) | 50 405(1) | 75 552(1) |
|---|---|---|---|
| Source[a] | $\delta\theta$ [ps] | $\delta\theta$ [ps] | $\delta\theta$ [ps] |
| $\Delta_I$ | 3.4 | 3.4 | 3.4 |
| DPO time base stability | 0.8 | 1.0 | 1.7 |
| Fit uncertainty | 1.5 | 1.5 | 1.0 |
| VOAs | 1.0 | 1.0 | 1.0 |
| PMD correction | 0.6 | 1.0 | 0.6 |
| Wavelength measurement | 0.2 | 0.5 | 0.7 |
| XCOR[b] interpolation | 0.3 | 0.3 | 0.3 |
| Estimate $n'''$ | 0.05 | 0.1 | 0.1 |
| SPM and XPM | <0.1 | <0.1 | <0.1 |
| Fast fiber length fluctuations | <0.1 | <0.1 | <0.1 |
| Uncertainty $L$[c] | <$10^{-4}$ | <$10^{-4}$ | <$10^{-4}$ |
| Total | 4.0 | 4.1 | 4.2 |

[a] For link configuration with FRMs.
[b] XCOR: cross correlation.
[c] Uncertainty of $L$ enters through the last term of Eq. (4).

The question arises whether the OWD determination, which involves the transmission of a PRBS rather than meaningful information, might introduce considerable amounts of dead time in the data transmission, thereby reducing the telecommunication capacity. This depends crucially on the rate of change of the OWD of a real-life fiber link. Recent time stability measurements performed on installed fiber loops of 60 km and 540 km length indicate that the TDEV increases to 10 ps after $8\times10^3$ s and $4\times10^3$ s, respectively [5,18]. This implies that in such installed fiber links, the OWD needs be re-calibrated only once per hour to maintain a time uncertainty smaller than 10 ps. Therefore, the dead time required for our method will be at most a few ms per hour, which is negligible. One may furthermore wonder whether the few-minute period, needed in the present setup for file transfer and cross correlation of captured DPO traces (Sec. 2.2), will somehow limit the achievable accuracy when used for TWTT between remote clocks. This appears not to be the case: a clock may be synchronized with ps stability (but with an unknown offset due the yet unknown OWD) by, for example, phase-locking to the same PRBS code as used for the OWD determination. After the OWD has been determined, and assuming the internal offset of the remote clock has not changed appreciably, the OWD is sent as a correction term to the remote clock which corrects its time offset accordingly (*e.g.* through an electronic variable delay line [18]). A particularly interesting option is to phase-lock the 10 Gb/s bit rate of the transmitter to the local reference clock, and use the recovered clock signal in the remote location as the oscillator of the 'slave clock'. Slow frequency variations of this oscillator due to fiber path length changes will be highly correlated with OWD variations. In this scenario, clock synchronization is not affected by the duration of the OWD determination (presently several minutes) as long as this duration is much shorter than the required time interval between successive OWD calibrations (once per hour to maintain <10 ps time uncertainty). Furthermore, it is worth noting that digital receivers based on field-programmable gate arrays (FPGAs) exist which can capture and process 10 Gb/s data in real time. Such digital signal processing may be used to implement the cross-correlation algorithm in real time (thus removing the need for time-consuming file transfer while speeding up the correlation itself), and is in-line with current trends in optical communication (coherent receivers). It may also present a cost-effective and compact alternative to the relatively expensive and voluminous DPO used in the present work.

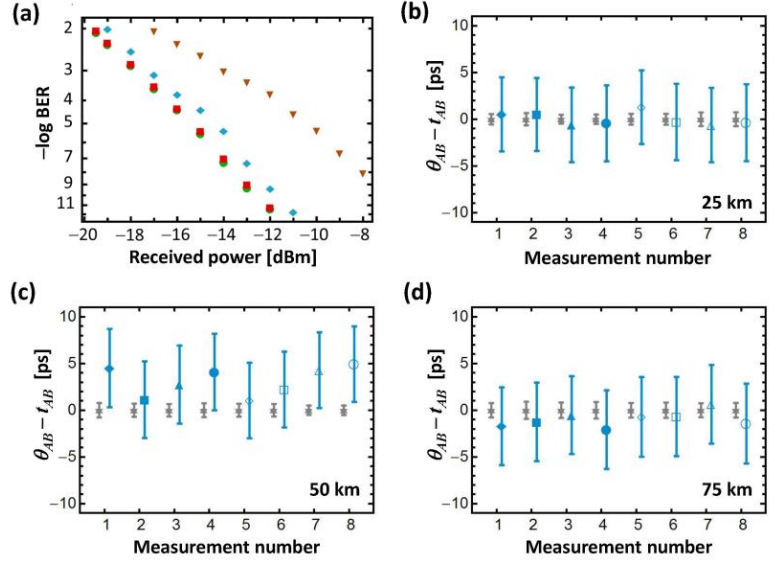

Fig. 4. Data transmission performance and accuracy of the OWD determination achieved using our method. (a) BER measurements for back-to-back (green dots), 25 km (red squares), 50 km (blue diamonds) and 75 km (brown triangles) fiber links. Error-free transmission (BER < $10^{-9}$) of 10 Gb/s data is achieved for the back-to-back, 25 km and 50 km links. For the 75 km link a BER slightly above $10^{-9}$ is achieved due to power limitations in the setup; however, the BER data show no sign of a noise floor, and with a slightly higher power level (or with dispersion compensation) a BER below $10^{-9}$ should be reached. (b) Delay differences $\theta_{AB} - t_{AB}$ (blue symbols) for the 25 km link. Gray stars represent the offset of the reference measurements $t_{AB}$ (which is zero by definition) and its measurement uncertainty. (c),(d) Same as (b), but for the 50 km and 75 km links, respectively. In (d), the symbols correspond directly to those in Fig. 3(a). For example, the first measurement value of $\theta_{AB} - t_{AB}$ (filled diamond) is found from combining the two filled-diamond values in Fig.3(a), which were obtained using different wavelength combinations and with the FRMs installed.

To our knowledge, the 4 ps measurement uncertainty reported here represents the most accurate long-haul signal propagation delay measurement to date, outperforming state-of-the-art satellite methods by at least one order of magnitude [2,3], as well as recently reported techniques for fiber-optical time transfer [5-7]. When implemented in a TWTT system, our method may also improve on the reach, communication bandwidth, and timing uncertainty of current state-of-the-art methods for simultaneous data and time transfer (sub-nanosecond and 1 Gb/s over up to 10 km distance) [12,13]. We furthermore note that the data-acquisition time required to achieve 4 ps measurement uncertainty is less than 1 ms, which is more than a thousand times faster than previous reported methods [5-7]. The combination of high time resolution and high stability results from the high bit rate of the 10 Gb/s data stream, and may improve further if higher bit rates are used. While no dispersion-compensation measures are taken in this experiment, we expect this will improve the bit-error-rate (BER) of the transmitted data. Indeed, Fig. 2(e) shows that after transmission through 75 km and 150 km of fiber, the eye diagram exhibits a progressive amount of pulse broadening due to dispersion. However, standard dispersion compensation methods (in combination with the optical amplifiers used here) will allow significantly longer distances to be covered, similar to typical commercially deployed optical data links. We furthermore note that the noise figure of SOAs is typically several dB larger than that of erbium-doped fiber amplifiers (EFDAs), which are commonly used in long-haul optical links. The results reported here therefore prove that the method works well also under sub-optimum conditions (using SOAs instead of EDFAs, and without dispersion compensation). A great advantage of SOAs is the possibility to amplify signals in a much wider wavelength range than the C-band in which EDFAs operate. A particular advantageous approach would be to use an in-band channel near the edge of the C-band for both data transmission and delay determination (making use of the EDFAs already present in the telecommunications system), while using out-of-band wavelength channels and SOAs for the return signals needed for cross correlation and delay determination. In this scenario, made possible by the use of different wavelengths and SOAs in our method, virtually no existing C-band capacity needs be sacrificed to transfer both data and time.

## 4. Conclusion and outlook

In conclusion, we report a method which enables delay measurements of 10 Gb/s optical data over 75 km distance with an unprecedented uncertainty of 4 ps. We expect that our method will be useful for TWTT and will find application in various areas of scientific research, including long-distance comparisons of atomic clocks and the realization of Coordinated Universal Time (UTC), fundamental physics research, geodesy, and radio astronomy through aperture synthesis. It is also worth noting that our method offers high-capacity data transmission along with a timing accuracy that is well sufficient to reliably synchronize (smart) electricity grids [20], the optical backbone of 4G mobile telecommunication networks [21], and networks for electronic financial transactions and trading, systems which have become crucially dependent on GNSS timing [1]. In general, our method may be used to extend existing optical communications infrastructure with a novel synchronization feature to mitigate the growing risk of jamming or outage of GNSS signals and the consequential failure of GNSS-dependent systems [1]. Moreover, a widely available optical source of picosecond timing may enable new techniques for secure communications, advanced beam-forming techniques for cooperative base stations for fast wireless Internet [21], and enhanced terrestrial positioning systems. For example, a grid of radio transmitters connected to the synchronized optical network could in principle enable positioning (in four-dimensional spacetime) with $c \times 4.0$ ps $\times \sqrt{4} = 2.4$ mm uncertainty. We therefore anticipate that the simultaneous transfer of data and time will become a standard feature of future optical networks, enabling accurate synchronization of next-generation high-capacity telecommunication networks and dependent network technologies, as well as hybrid optical-wireless systems providing terrestrial navigation and positioning with unprecedented accuracy.

## Acknowledgments

This work was supported through the SURFnet Gigaport 3 program. J.C.J.K. thanks the Netherlands Organisation for Scientific Research (NWO) and the Dutch Technology Foundation (STW) for support. Henk Peek is acknowledged for suggesting dispersion reconstruction by use of a third wavelength, Erwin Bente for loan of the wavelength meter, and Kjeld Eikema and Tjeerd Pinkert for loan of and assistance with the rubidium atomic clock.